\documentclass[runningheads]{llncs}
\usepackage[T1]{fontenc}
\usepackage{graphicx}
\usepackage{amsmath} 
\usepackage{amsmath}
\usepackage{amssymb}
\newcommand{\thename}{PixCrypt}

\usepackage{marvosym}

\newcommand{\equalcontrib}{\textsuperscript{*}}
\newcommand{\corresponding}{\textsuperscript{\Letter}}

\usepackage{booktabs}

\newcommand{\Enc}{\mathrm{Enc}}
\newcommand{\pt}{\mathrm{pt}}
\usepackage{multirow} 
\usepackage{makecell}
\usepackage{microtype}

\usepackage[ruled,vlined]{algorithm2e}
\usepackage{accessibility}

\usepackage[colorlinks=true,
            linkcolor=blue,
            citecolor=blue,
            urlcolor=blue]{hyperref}

\begin{document}
%
\title{PixCrypt: Fast Fine-Grained FHE with Range-Aware Caching}
%
%



\author{
Chao Wang\inst{1}\equalcontrib \and
Shubing Yang\inst{2}\equalcontrib \and
Xiaoyan Sun\inst{1} \and
Yan Bai\inst{2} \and
Jun Dai\inst{1}\corresponding \and
Dongfang Zhao\inst{2}\corresponding
}

\authorrunning{C. Wang et al.}

\institute{
Worcester Polytechnic Institute, Worcester, MA 01609, USA\\
\email{\{cwang17,xsun7,jdai\}@wpi.edu}
\and
University of Washington, Seattle, WA 98195, USA\\
\email{\{sueyoung,yanb\}@uw.edu, dzhao@cs.washington.edu}
}
%
%
\maketitle              
\begingroup
\renewcommand{\thefootnote}{*}
\footnotetext{Chao Wang and Shubing Yang contributed equally to this work.}

\renewcommand{\thefootnote}{\Letter}
\footnotetext{Jun Dai and Dongfang Zhao are the corresponding authors.}
\endgroup

\begingroup
\renewcommand{\thefootnote}{}
\footnotetext{This paper has been accepted at the 28th International Conference on Information and Communications Security (ICICS 2026).}
\endgroup

\begin{abstract}
Many analytics tasks require secure computation over encrypted data.
In particular, fine-grained data such as pixel-level images require higher precision, as every pixel can directly affect outcomes in tasks like tumor segmentation and anomaly detection. While Multi-Party Computation (MPC) is interactive, Differential Privacy (DP) protects only aggregate values, and Partially Homomorphic Encryption (PHE) lacks multiplicative support, none of them can efficiently handle fine-grained data analytics. Fully Homomorphic Encryption (FHE) uniquely enables arbitrary operations on encrypted pixels but remains computationally expensive, posing significant challenges for both software and hardware accelerators. We present PixCrypt, a caching-based acceleration mechanism for fine-grained fully homomorphic encryption. PixCrypt replaces expensive fresh ciphertext generation with cache retrieval and coefficient-level operations across CKKS, BFV, and BGV, while randomized reconstruction ensures that ciphertexts do not repeat. Its linear noise growth reduces the need for bootstrapping and lowers NTT load, improving hardware accelerator efficiency. This design yields up to 35× faster fine-grained encryption and maintains IND-CPA (Indistinguishability under Chosen Plaintext Attack) security. Experiments on five real-world pixel-level image processing tasks show that PixCrypt significantly improves the practicality of FHE for privacy-preserving analytics.

\end{abstract}
\section{Introduction}

The confidentiality of medical data, particularly diagnostic images stored in healthcare databases, faces severe risks under escalating cyberattacks. High-profile incidents such as the 2024 data leak and the 2023 ransomware attack on Lehigh Valley Health Network expose critical vulnerabilities in existing database and imaging management systems. Many encryption methods reduce computation by using coarse processing, but this sacrifices pixel-level access. In medical imaging, \textbf{fine-grained} operations are essential for tasks like noise reduction, anomaly detection, and spatial analysis. Without this granularity, system effectiveness and clinical value are both compromised.

Existing secure image analysis techniques have key limitations: MPC requires distributed coordination~\cite{mpc2019}, DP adds nondeterministic noise~\cite{dp2008}, and PHE lacks full computational expressiveness~\cite{he_survey}. Fully Homomorphic Encryption (FHE), by contrast, offers a non-interactive design, supports diverse computation types, and provides strong semantic security, making it well suited for encrypted fine-grained data analysis.

Despite its expressiveness, FHE remains difficult to deploy in practice due to high computational cost. Naive pixel-level encryption and evaluation introduce prohibitive latency and memory usage, especially for high-resolution images or iterative operations. These constraints make direct FHE solutions impractical for real-time or resource-limited medical imaging, motivating the need for more efficient and scalable methods. We focus on CKKS, BFV, and BGV due to their suitability for fine-grained arithmetic, while excluding TFHE, which is optimized for Boolean circuits and does not natively support floating-point arithmetic required for common image processing operations~\cite{torusCircuit}.

To address the substantial performance bottleneck of fine-grained FHE in encrypted image processing, we propose PixCrypt, an optimization protocol that leverages caching mechanisms to significantly accelerate encrypted computations while maintaining fine-grained operational granularity. PixCrypt adopts two specialized caching strategies: square-root-based using a hybrid decomposition (multiplication and addition) based on square roots and parity-based using an additive decomposition into even components plus a parity bit. By reusing these pre-encrypted values during pixel-wise computation, PixCrypt significantly reduces redundant encryption operations and improves overall processing efficiency.

We implement PixCrypt on CKKS, BFV, and BGV, and evaluate it on the USC-SIPI~\cite{USCSIPI} dataset across five pixel-level tasks: mean filtering, image matching, ciphertext watermarking, segmentation, and binary classification. PixCrypt delivers up to 35× faster per image encryption and preserves IND-CPA security. The key insights and contributions are summarized as follows:
\begin{itemize} 
     \item We propose a caching-based protocol that accelerates homomorphic encryption for fine-grained data through square-root-based and parity-based caching, which offer different trade-offs between offline precomputation and online encryption, while our controlled-noise design further reduces the reliance on costly bootstrapping.
     (Section~\ref{cachemethodology} and~\ref{bootstrapping})

    \item We design a randomization layer that adds noise during ciphertext reconstruction to ensure IND-CPA security while enabling multiple computations on the same encrypted data (Section~\ref{Randomization Mechanism}). We formally prove that PixCrypt's IND-CPA security is guaranteed by the security of the underlying homomorphic encryption schemes. (Section~\ref{INDCPA})
    
    \item We conduct extensive evaluations on four distinct types of image data and five real-world pixel-level tasks and demonstrate substantial improvements in encryption efficiency across diverse scenarios, such as filtering, watermarking, segmentation and classification. (Section~\ref{casestudy})
\end{itemize}

\section{Related Work and Preliminaries}

\subsection{FHE and Acceleration}
\label{relatedwork: FHE}

Fully Homomorphic Encryption (FHE) enables computation on encrypted data without requiring decryption, a concept first proposed by Rivest~\cite{rivest1978data} and practically realized by Gentry~\cite{gentry2009}. Subsequent research has focused on enhancing its efficiency and usability~\cite{blatt2020optimized}, leading to several widely used schemes including BGV~\cite{BGV2014}, BFV~\cite{FV2012,brakerski2012}, CKKS~\cite{ckks2016}, and TFHE~\cite{torusCircuit}. These schemes differ in their arithmetic support and application contexts: TFHE excels in boolean circuits with fast bootstrapping~\cite{ccsTFHEAccelerator}, CKKS supports approximate real/complex arithmetic for ML tasks, and BFV/BGV focus on exact integer operations. While they share similar parameter foundations~\cite{HEStandard}, each offers trade-offs in circuit depth, data types, and performance. Notably, BGV handles deeper circuits without bootstrapping, and BFV achieves high efficiency in certain tasks, further boosted by optimizations such as RNS techniques~\cite{improveBFV2019} and the BASALISC accelerator~\cite{geelen2022basalisc}.

CKKS has particular traction in privacy-preserving machine learning~\cite{boemer2020mp2ml} for its support of approximate arithmetic over $\mathbb{C}^{N/2}$ via batching in $\mathbb{Z}_Q[X]/(X^N + 1)$. This scheme balances precision and performance, with parameters like ring dimension $N$ and modulus $Q$ (Table~\ref{table:parameters}) critically affecting efficiency and security~\cite{bts2022}.

\begin{table}[h]
\setlength{\abovecaptionskip}{2pt}
\small
\caption{Parameters and their definitions in the CKKS scheme}
\centering
\begin{tabular}{c l}
\hline
\textbf{Symbol} & \textbf{Definition} \\ \hline
$N$ & The ring dimension\\ 
$Q$ & The ciphertext modulus\\ 
$P$ & The auxiliary modulus (for relinearization)\\ 
$L$ & Maximum level (multiplicative)  \\ 
$\ell$ & Current (multiplicative) level of a ciphertext \\ 
$\text{evk}_{\text{mul}}$ & Evaluation key (evk) for Multiplications \\ 
$\Delta$ &Scaling factor of a CKKS plaintext\\
$\lambda$ & Security parameter of a given CKKS instance \\ \hline
\end{tabular}

\label{table:parameters}
\end{table}

To improve the performance of homomorphic encryption, various acceleration techniques have been proposed. These include memory-aware bootstrapping for CKKS~\cite{agrawal2023mad,agrawal2023fab}, gadget decomposition for optimizing bottleneck operations~\cite{gadgetDecomposition}, and hardware-level enhancements such as the reconfigurable FFT/NTT hardware~\cite{ABCFHE}, FHE for CGRA~\cite{DAC24}, Meta-OP unified operator~\cite{DAC24.2}, accelerator for fast matrix-vector product~\cite{dac23}, accelerator for TFHE~\cite{dac22}, accelerating
FHE with processing in-memory~\cite{dac21} and CUDA-optimized libraries~\cite{phantom}. Beyond FHE, efforts to speed up Partially Homomorphic Encryption (PHE) have also progressed. For instance, Rache~\cite{sigmod2023} uses parallel caching to accelerate PHE for outsourced cloud databases. However, its scope is limited to PHE schemes like Paillier~\cite{paillier1999} and Symmetria~\cite{symmetria}.

\subsection{Provable Security}

To reason about the security of a cryptographic scheme, we specify the \textit{security goal}, \textit{threat model}, and \textit{assumptions}. The security goal states what the scheme must protect; the threat model defines the adversary’s capabilities; and assumptions describe the computational limits and primitives underlying the scheme.

We adopt \textit{indistinguishability under chosen-plaintext attack} (IND-CPA). The adversary may request polynomially many encryptions of chosen plaintexts and then submit two equal-length challenge messages $m_0$ and $m_1$. The challenger samples a uniformly random bit $b \leftarrow \{0,1\}$ and fresh encryption randomness $\rho$, and returns the challenge ciphertext $c^* \leftarrow \mathsf{Enc}_{pk}(m_b;\rho)$. After receiving $c^*$, the adversary outputs a bit $b'$. The adversary's IND-CPA advantage is formally defined as
\[
\mathsf{Adv}^{\mathsf{IND\text{-}CPA}}_{\Pi,\mathcal{A}}(\lambda)
=
\left|
\Pr[b'=b]-\frac{1}{2}
\right|.
\]
A scheme $\Pi$ is IND-CPA secure if $\mathsf{Adv}^{\mathsf{IND\text{-}CPA}}_{\Pi,\mathcal{A}}(\lambda)$ is negligible in the security parameter $\lambda$ for every probabilistic polynomial-time adversary $\mathcal{A}$. A function $\mu(\lambda)$ is negligible if, for every positive polynomial $p(\lambda)$, there exists $\lambda_0$ such that $\mu(\lambda) < 1/p(\lambda)$ for all $\lambda \geq \lambda_0$. We use the standard fact that a polynomially bounded sum of negligible functions remains negligible. This fact forms the basis for proving that our construction satisfies IND-CPA security in Section~\ref{INDCPA}.

\begin{figure*}[htbp]
    \centering
    \includegraphics[width=1\textwidth]{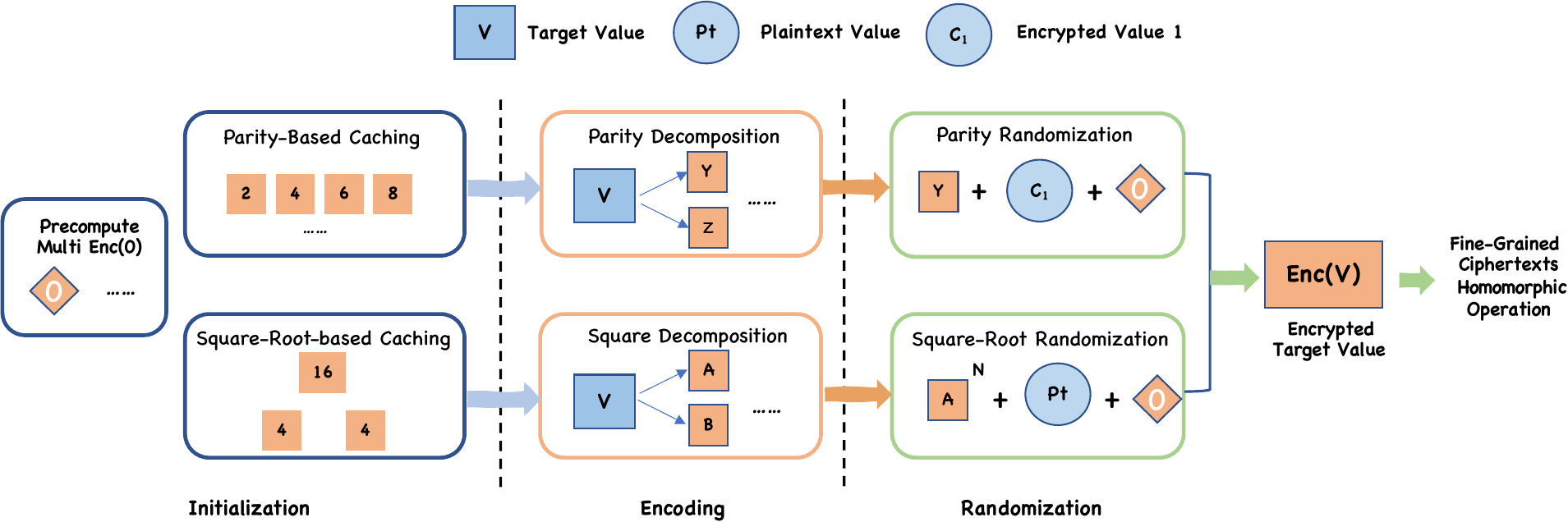}
    \caption{
    Overview of \thename. Parity-based and square-root-based caching accelerate scalar homomorphic encryption by combining plaintext with cached ciphertexts and injecting randomized noise to construct new ciphertexts, while preserving IND-CPA security. Blue denotes plaintext; orange denotes ciphertext.
    }
    \label{fig:design}
\end{figure*}

\section{Proposed Method}

Our approach accelerates fine-grained data encryption under fully homomorphic encryption (FHE). While schemes such as CKKS, BFV, and BGV enable encrypted-domain computation, they incur substantial cost when encrypting individual data elements independently. To mitigate this bottleneck, we introduce two caching strategies that reduce each element encryption cost (Figure~\ref{fig:design}). 

The design of caching layer is presented in Section~\ref{cachemethodology}. The randomization layer is described in Section~\ref{Randomization Mechanism}, and the compliance of our approach with IND-CPA security is discussed in Section~\ref{INDCPA}.

\subsection{Caching Layer}
\label{cachemethodology}
Many real-world applications require fine-grained FHE operations on data with bounded integer values, such as sensor readings, rating scores, categorical variables, and digital images. A key observation is that these values are drawn from a small, finite domain (e.g., $[0, 255]$ for 8-bit data), and encryption is repeatedly applied to individual elements from this limited set. We leverage this bounded and discrete structure to accelerate encryption by precomputing and caching a compact set of ``basis'' ciphertexts offline, then synthesizing encryptions of arbitrary values through lightweight homomorphic operations on these cached elements. This approach trades off between cache size and online computation cost: larger caches allow purely additive synthesis with minimal computation, whereas smaller caches require extra multiplicative steps but significantly reduce storage overhead. We study two instantiations of this idea: one using an additive decomposition into even components plus a parity bit, and another using a multiplicative–additive decomposition based on square roots.

\paragraph{\textbf{Parity-based Caching}}
This method adopts an \emph{additive decomposition} viewpoint.
Let $\mathcal{D}=\{0,1,\dots,M\}$ be a bounded integer domain (e.g., $M=255$ for 8-bit pixels). Fix a public-key FHE scheme $\Pi=(\mathrm{Setup},\mathrm{KeyGen},\mathrm{Enc},\mathrm{Dec},\mathrm{Eval})$ that supports ciphertext--ciphertext addition and either ciphertext--plaintext addition or, alternatively, ciphertext--ciphertext addition with cached small constants.
We write $\mathrm{pt}(\cdot)$ for the scheme's plaintext encoding (with a fixed slot layout and, for CKKS, a fixed scale), and use $\oplus,\ominus$ for homomorphic addition and subtraction.
We precompute only \emph{even} encryptions together with an encryption of 1 and a small pool of fresh zeros: $\mathcal{C}_{\mathrm{even}} = \{\mathrm{Enc}(0), \mathrm{Enc}(2), \dots, \mathrm{Enc}(2\lfloor M/2\rfloor)\}$, $\mathcal{C}_1 = \mathrm{Enc}(1)$, and $\mathcal{Z} = \{\mathrm{Enc}_{\mathrm{rand}}(0)_1, \dots, \mathrm{Enc}_{\mathrm{rand}}(0)_{N_z}\}$, where $\mathcal{Z}$ denotes a shared pool of $N_z$ fresh, independent encryptions of zero used for re-randomization in both caching strategies, and $N_z$ is a system parameter controlling the pool size. For any plaintext $t \in \mathcal{D}$, we decompose it using parity as $t = 2v + \delta$ where $v = \lfloor t/2 \rfloor$ and $\delta = t - 2v \in \{0,1\}$. We then construct the encryption by retrieving $\mathrm{Enc}(2v)$ from $\mathcal{C}_{\mathrm{even}}$ and sampling a fresh $\mathrm{Enc}(0)$ from $\mathcal{Z}$ for randomization.\par
The output of Algorithm~\ref{alg:parity_caching} serves as the input to Algorithm~\ref{alg:parity_randomization}.

\begin{algorithm}[t]
\normalsize
\caption{Parity-based Caching}
\label{alg:parity_caching}
\KwIn{Value Collection $\mathcal{D}=\{0,1,\dots,M\}$; FHE scheme $\Pi=(\mathrm{Setup},\mathrm{KeyGen},\mathrm{Enc},\mathrm{Dec},\mathrm{Eval})$; Plaintext $t \in \mathcal{D}$}
\KwOut{$CachedList$}
\tcp{Precomputation phase (done once)}
\If{cache not initialized}{
    $\mathcal{C}_{\mathrm{even}} \leftarrow \{\mathrm{Enc}(0), \mathrm{Enc}(2), \mathrm{Enc}(4), \dots, \mathrm{Enc}(2\lfloor M/2\rfloor)\}$\;
    $\mathcal{C}_1 \leftarrow \mathrm{Enc}(1)$ \tcp{reusable encryption of 1}
    $\mathcal{Z} \leftarrow \{\mathrm{Enc}_{\mathrm{rand}}(0)_1, \mathrm{Enc}_{\mathrm{rand}}(0)_2, \dots, \mathrm{Enc}_{\mathrm{rand}}(0)_{N_z}\}$ \tcp{shared zero pool for re-randomization}
}
\tcp{Parity decomposition}
$v \leftarrow \lfloor t/2 \rfloor$\;
$\delta \leftarrow t - 2v$ \tcp{$\delta \in \{0,1\}$}
\tcp{Retrieve base even encryption}
$C_{2v} \leftarrow$ lookup $\mathrm{Enc}(2v)$ from $\mathcal{C}_{\mathrm{even}}$\;
\Return $ \{\mathcal{C}_{\mathrm{even}}, \mathcal{C}_1, \mathcal{Z}\}$;
\end{algorithm}

\paragraph{\textbf{Square-Root-based Caching}}
In contrast, this method uses a \emph{multiplicative--additive hybrid} decomposition.
Let $\mathcal{D}=\{0,1,\dots,M\}$ be a bounded integer domain (e.g., $M=255$ for 8-bit pixels).
Fix a public-key FHE scheme $\Pi=(\mathrm{Setup},\mathrm{KeyGen},\mathrm{Enc},\mathrm{Dec},\mathrm{Eval})$ that supports ciphertext--ciphertext multiplication, ciphertext--plaintext multiplication, and ciphertext--plaintext addition.
We write $\mathrm{pt}(\cdot)$ for the scheme's plaintext encoding (with a fixed slot layout and, for CKKS, a fixed scale), and use $\oplus,\ominus,\odot$ for homomorphic addition, subtraction and multiplication.
We precompute only \emph{square root} encryptions together with a small pool of $N_z$ fresh zeros: $\mathcal{C}_{\mathrm{sqrt}} = \{\mathrm{Enc}(0), \mathrm{Enc}(1), \dots, \mathrm{Enc}(\lfloor\sqrt{M}\rfloor)\}$ and $\mathcal{Z} = \{\mathrm{Enc}_{\mathrm{rand}}(0)_1, \dots, \mathrm{Enc}_{\mathrm{rand}}(0)_{N_z}\}$.
Additionally, we precompute plaintext encodings for all square root values and remainders: $\mathcal{P}_{\mathrm{sqrt}} = \{\mathrm{pt}(0), \mathrm{pt}(1), \dots, \mathrm{pt}(\lfloor\sqrt{M}\rfloor)\}$ and $\mathcal{P}_{\mathrm{rem}} = \{\mathrm{pt}(0), \mathrm{pt}(1), \dots, \mathrm{pt}(2\lfloor\sqrt{M}\rfloor)\}$.
For any $v\in\mathcal{D}$, express the square decomposition $v = k^2 + r$, where $k=\lfloor\sqrt{v}\rfloor$ and $r=v-k^2\in\{0,1,\dots,2\lfloor\sqrt{M}\rfloor\}$.
To synthesize an encryption of $v$, retrieve $\mathrm{Enc}(k)\in\mathcal{C}_{\mathrm{sqrt}}$ and set
\begin{equation}
C_v =
\bigl(\mathrm{Enc}(k)\odot\mathrm{pt}(k)\bigr)
\oplus\mathrm{pt}(r)\oplus Z,
\qquad
Z\gets\mathcal{Z}.\mathrm{pop}(),
\label{eq:sqrt_synth}
\end{equation}
where the multiplication $\mathrm{Enc}(k)\odot\mathrm{pt}(k)$ computes $k^2$ homomorphically, followed by addition of the remainder $r$ and fresh zero re-randomization.\par
The output of Algorithm~\ref{alg:square_based_caching} serves as the input to Algorithm~\ref{alg:sr_construct}.

\begin{algorithm}[t]
\normalsize
\caption{Square-Root-based Caching}
\label{alg:square_based_caching}
\KwIn{Domain $\mathcal{D}=\{0,1,\dots,M\}$; FHE scheme $\Pi=(\mathrm{Setup},\mathrm{KeyGen},\mathrm{Enc},\mathrm{Dec},\mathrm{Eval})$; Plaintext $t \in \mathcal{D}$}
\KwOut{$\textit{CachedList}$}
Initialize HE context for $\Pi$ (parameters, keys, encoder/decoder, evaluator)\;
\textbf{Build caches}\;
\For{$k=0$ \KwTo $\lfloor\sqrt{M}\rfloor$}{
$\mathcal{C}_{\mathrm{sqrt}}[k] \leftarrow \mathrm{Enc}(k)$ \tcp*[r]{ciphertexts of $0,\dots,\lfloor\sqrt{M}\rfloor$}
}
\For{$k=0$ \KwTo $\lfloor\sqrt{M}\rfloor$}{
$\mathcal{P}_{\mathrm{sqrt}}[k] \leftarrow \mathrm{pt}(k)$ \tcp*[r]{plaintext encodings}
}
\For{$r=0$ \KwTo $2\lfloor\sqrt{M}\rfloor$}{
$\mathcal{P}_{\mathrm{rem}}[r] \leftarrow \mathrm{pt}(r)$ \tcp*[r]{plaintext remainders}
}
\For{$i=1$ \KwTo $N_z$}{
$\mathcal{Z}[i] \leftarrow \mathrm{Enc}_{\mathrm{rand}}(0)$ \tcp*[r]{independent encryptions of $0$}
}
\Return $\textit{CachedList}=\{\mathcal{C}_{\mathrm{sqrt}},\mathcal{P}_{\mathrm{sqrt}},\mathcal{P}_{\mathrm{rem}},\mathcal{Z}\}$\;
\end{algorithm}

\subsection{Randomization Layer}
\label{Randomization Mechanism}

If ciphertexts are exposed and attackers have prior knowledge about plaintext distributions or access to an encryption oracle, encryption schemes become susceptible to chosen-plaintext attacks (CPA). Thus, it is essential to introduce randomness each time a cached ciphertext is used.

\begin{algorithm}[!tbp]
\small
\caption{Parity-Based Randomization}
\label{alg:parity_randomization}

\KwIn{Plaintext $t \in \{0,1,\dots,M\}$; Cache
$\mathcal{C}_{\mathrm{even}}, \mathcal{C}_1$; Zero pool
$\mathcal{Z} =
\{\mathrm{Enc}_{\mathrm{rand}}(0)_1,\dots,
\mathrm{Enc}_{\mathrm{rand}}(0)_{N_z}\}$.}

\KwOut{Randomized ciphertext $C'$ such that
$\mathsf{Dec}(C') = t$.}

\BlankLine

$v \gets \lfloor t/2 \rfloor$\;
$\delta \gets t - 2v$\;

\BlankLine

$C_{\mathrm{base}} \gets
\mathcal{C}_{\mathrm{even}}[v]$\;

$Z \gets \mathcal{Z}.\mathit{pop}()$\;

\BlankLine

\uIf{$\delta = 0$}{
    $C' \gets C_{\mathrm{base}} \oplus Z$\;
}
\Else{
    $C' \gets C_{\mathrm{base}} \oplus
    \mathcal{C}_1 \oplus Z$\;
}

\BlankLine

\Return $C'$\;

\end{algorithm}

\begin{algorithm}[!tbp]
\small
\caption{Square-Root-Based Randomization}
\label{alg:sr_construct}
\KwIn{
  Domain $\mathcal{D} = \{0,1,\dots,M\}$; \\
  Target plaintext $t \in \mathcal{D}$; \\
  Ciphertext cache $\mathcal{C}_{\mathrm{sqrt}} = \{\mathsf{Enc}(k)\}_{k=0}^{\lfloor \sqrt{M} \rfloor}$; \\
  Plaintext tables: \\
  \quad $\mathcal{P}_{\mathrm{sqrt}} = \{\mathsf{pt}(k)\}_{k=0}^{\lfloor \sqrt{M} \rfloor}$, \\
  \quad $\mathcal{P}_{\mathrm{rem}} = \{\mathsf{pt}(r)\}_{r=0}^{2\lfloor \sqrt{M} \rfloor}$; \\
  Zero pool $\mathcal{Z} = \{\mathrm{Enc}_{\mathrm{rand}}(0)_1, \dots, \mathrm{Enc}_{\mathrm{rand}}(0)_{N_z}\}$.
}
\KwOut{Ciphertext $C_t$ such that $\mathsf{Dec}(C_t) = t$.}
\BlankLine
$k \gets \lfloor \sqrt{t} \rfloor$\;
$r \gets t - k^2$\
\BlankLine
$C_k \gets \mathcal{C}_{\mathrm{sqrt}}[k]$\ 
$P_k \gets \mathcal{P}_{\mathrm{sqrt}}[k]$\ 
$B_{\text{base}} \gets C_k \odot P_k$\
\BlankLine
\If{\textnormal{scheme is CKKS}}{
  $\mathsf{RescaleTo}(B_{\text{base}},\ \texttt{target\_scale})$\;
}
\BlankLine
$P_r \gets \mathcal{P}_{\mathrm{rem}}[r]$\ 
$B_{\text{sum}} \gets B_{\text{base}} \oplus P_r$\;
\BlankLine
$Z \gets \mathcal{Z}.\mathit{pop}()$\;
$C_t \gets B_{\text{sum}} \oplus Z$\;
\BlankLine
\Return $C_t$\;
\end{algorithm}

\paragraph{\textbf{Parity-based Randomization}}
Given a plaintext value $t$, we decompose it based on its parity as $t = 2v + \delta$ where $v = \lfloor t/2 \rfloor$ and $\delta \in \{0,1\}$ represents the remainder (i.e., whether $t$ is even or odd). 

To generate a fresh ciphertext encrypting the same value $t$, we reconstruct it from cached components: precomputed even ciphertexts from cache $\mathcal{C}_{\mathrm{even}}$, a unit ciphertext $\mathcal{C}_1 = \mathrm{Enc}(1)$, and fresh zero ciphertexts from cache $\mathcal{Z}$. The reconstruction process depends on the parity bit $\delta$:
\begin{align*}
C' \leftarrow \begin{cases}
\mathrm{Enc}(2v) \oplus Z, & \text{if } \delta = 0 \text{ (even)}, \\
\mathrm{Enc}(2v) \oplus \mathcal{C}_1 \oplus Z, & \text{if } \delta = 1 \text{ (odd)},
\end{cases}
\end{align*}
where $Z$ is a freshly sampled encryption of zero from $\mathcal{Z}$. A fresh zero is popped from the precomputed pool for each reconstruction. This approach preserves the plaintext value while introducing fresh encryption randomness through the zero ciphertext, as shown in Algorithm~\ref{alg:parity_randomization}. The method leverages the additive homomorphic property of the encryption scheme to combine cached ciphertexts with fresh randomness efficiently.

\paragraph{\textbf{Square-Root-based Randomization.}}
Given a target value $t \in \mathcal{D} = \{0, \ldots, M\}$ and precomputed
resources including a square root cache
$\mathcal{C}_{\mathrm{sqrt}} = \{\Enc(0),\allowbreak \Enc(1),\allowbreak \ldots,\allowbreak \Enc(\lfloor\sqrt{M}\rfloor)\}$,
plaintext tables
$\mathcal{P}_{\mathrm{sqrt}} = \{\pt(0), \ldots, \pt(\lfloor\sqrt{M}\rfloor)\}$ and
$\mathcal{P}_{\mathrm{rem}} = \{\pt(0), \ldots, \pt(2\lfloor\sqrt{M}\rfloor)\}$, and a
zero-pool
$\mathcal{Z} = \{\mathrm{Enc}_{\mathrm{rand}}(0)_1, \dots, \mathrm{Enc}_{\mathrm{rand}}(0)_{N_z}\}$,
we construct a fresh ciphertext encrypting $t$ through square-root decomposition.
As shown in Algorithm~\ref{alg:sr_construct}, the construction begins by
decomposing the target as $t = k^2 + r$ where $k = \lfloor\sqrt{t}\rfloor$ and
$r = t - k^2$. We then compute $k^2$ homomorphically using the precomputed
encryption and plaintext values as $B \leftarrow \mathcal{C}_{\mathrm{sqrt}}[k] \odot \mathcal{P}_{\mathrm{sqrt}}[k]$.
For CKKS schemes, we perform scale management by rescaling $B$ to the target scale before proceeding.
Next, we incorporate the remainder through plaintext addition, $C_t \leftarrow B \oplus \mathcal{P}_{\mathrm{rem}}[r]$.
Finally, we achieve re-randomization by sampling a fresh zero encryption $Z$ from the zero-pool $\mathcal{Z}$ and adding it to the result, $C_t \leftarrow C_t \oplus Z$.
This approach efficiently constructs target ciphertexts by leveraging precomputed square values up to $\sqrt{M}$, while ensuring cryptographic freshness through the zero-pool mechanism. The square-root decomposition reduces the computational overhead compared to direct homomorphic arithmetic while maintaining the desired plaintext value.

\noindent\textbf{Example.}
To encrypt $t=15$ where $\mathcal{D}=\{0,\dots,16\}$: \textbf{Parity-based}: Decompose $15 = 2 \cdot 7 + 1$, then compute $\mathrm{Enc}(15) = \mathrm{Enc}(14) \oplus \mathrm{Enc}(1) \oplus Z$ using only additions. \textbf{Square-root}: Decompose $15 = 3^2 + 6$, then compute $\mathrm{Enc}(15) = (\mathrm{Enc}(3) \odot \mathrm{pt}(3)) \oplus \mathrm{pt}(6) \oplus Z$ using multiplication and addition. Here $Z$ is a fresh $\mathrm{Enc}(0)$ for randomization and $\mathrm{pt}(\cdot)$ denotes plaintext encoding.

\subsection{Bootstrapping Reduction via Plaintext-Oriented Encoding}
\label{bootstrapping}
Bootstrapping is essential but costly in FHE, triggered when ciphertext noise exceeds a threshold. We propose a strategy that fundamentally reduces noise growth by shifting arithmetic computation to the plaintext domain. Instead of using ciphertext-ciphertext multiplications, our method combines cached ciphertexts with low-noise plaintext operations, keeping noise growth strictly additive. The key advantage is that while our approach introduces additional noise, this growth remains predictable and tightly bounded—linear (pt--ct) rather than multiplicative (ct--ct) noise accumulation in traditional methods. 

Let $d$ denote the operational depth, i.e., the maximum number of sequential homomorphic operations supported before bootstrapping is required, and $O(\cdot)$ denote standard asymptotic (big-O) notation. Let $\mathsf{Enc}_{pk}(m)$ denote the encryption with initial noise level $\sigma_0$, where $t$ denotes the number of sequential homomorphic operations. $\sigma_0$ is the initial noise level of a fresh ciphertext, $\Delta_{\mathrm{add}}$ and $\Delta_{\mathrm{mult}}$ denote the per-operation noise increments for plaintext-ciphertext addition and ciphertext-ciphertext multiplication respectively, $\epsilon_j$ is the noise perturbation at step $j$, and $\mathsf{threshold}$ denotes the maximum noise level before bootstrapping is triggered. The noise evolution follows:

\begin{align*}
\mathsf{noise}_{\mathrm{ours}}(t)
&= \sigma_0 + \sum_{j=1}^{t} \Delta_{\mathrm{add},j}
   + \underbrace{O(\log M)}_{\text{controlled overhead}} \\ 
&= O(\sigma_0 + t \cdot \Delta_{\mathrm{add}} + \log M), \\[4pt] 
\mathsf{noise}_{\mathrm{baseline}}(t)
&= \sigma_0 \cdot \prod_{j=1}^{t} (1 + \epsilon_j)
   \;\le\; O(\sigma_0 \cdot \Delta_{\mathrm{mult}}^{\,t}). 
\end{align*}

Critically, our noise growth is linear in $t$ and bounded by
$O(t \cdot \Delta_{\mathrm{add}} + \log M)$. 
Our method's additive noise accumulation, deterministically bounded by
the worst-case pt--ct operation noise $\Delta_{\mathrm{add}}$, permits precise
noise-budget management, unlike the multiplicative growth in baseline approaches. This controlled noise extends operational depth before bootstrapping, i.e., $d_{\mathrm{ours}} \gg d_{\mathrm{baseline}}$, specifically as follows:
\begin{align*}
d_{\mathrm{ours}} &= O\!\left(\frac{\mathsf{threshold}}{\Delta_{\mathrm{add}}}\right) 
\gg d_{\mathrm{baseline}} = O\!\left(\log_{\Delta_{\mathrm{mult}}} \mathsf{threshold}\right). 
\end{align*}

\begin{table*}[t]
\setlength{\abovecaptionskip}{2pt}
\centering
\small
\caption{Comparison of complexity in terms of precomputation time, encryption time, and space.}
\label{tab:complexity_comparison}

\resizebox{\textwidth}{!}{%
\begin{tabular}{lcccl}
\toprule
\textbf{Method} & \textbf{Scheme} & \textbf{Precomputation Time} & \textbf{Encryption Time} & \textbf{Space Complexity} \\
\midrule

Without Caching (Baseline) 
& CKKS / BFV / BGV 
& None 
& $O(n \cdot N \log N)$ 
& None \\

Rache~\cite{sigmod2023} (Baseline) 
& CKKS / BFV / BGV 
& $O(\lceil \log_r M \rceil \cdot t_e)$ 
& $O(n \cdot \lceil \log_r M \rceil)$ 
& $O(\lceil \log_r M \rceil)$ \\

Parity-based (\textbf{Ours}) 
& CKKS / BFV / BGV 
& $O(M/2 \cdot t_e)$ 
& $O(n)$ 
& $O(M/2)$ \\

Square-root-based (\textbf{Ours}) 
& CKKS / BFV / BGV 
& $O(\sqrt{M} \cdot t_e)$ 
& $O(n \cdot \log \sqrt{M})$ 
& $O(\sqrt{M})$ \\

\bottomrule
\end{tabular}%
}

\vspace{0.5em}

\begin{minipage}{\textwidth}
\scriptsize
\textit{
Note: 
$n$ = number of pixels, 
$M$ = domain size (e.g., 255 for 8-bit), 
$N$ = polynomial modulus degree, 
$t_e$ = time complexity of single encryption ($O(N \log N)$), 
$r$ = radix base.
}
\end{minipage}
\end{table*}

\subsection{Complexity Analysis}
\label{sec:complexity_analysis}

We compare the computational and space complexity of the proposed caching strategies with the baselines. Let $n$ denote the number of pixels, $M$ the domain size (typically $255$ for 8-bit data), and $t_e$ the cost of one encryption, typically $O(N\log N)$ for polynomial modulus degree $N$. Without caching, encryption costs $O(n\cdot N\log N)$, whereas Rache uses $O(\lceil\log_r M\rceil)$ space and $O(n\cdot\lceil\log_r M\rceil)$ online time. Our parity-based method uses $O(M/2)$ space and $O(n)$ online time, while the square-root-based method uses $O(\sqrt{M})$ space with homomorphic multiplication during online synthesis. Table~\ref{tab:complexity_comparison} summarizes these trade-offs. We instantiate these space bounds for 8, 12, and 16-bit domains in Section~\ref{sec:bitdepth}.

\subsection{Security Analysis and Proof}
\label{INDCPA}

In this section, we demonstrate that the modified FHE schemes, incorporating parity-based and square-root-based caching strategies, maintain IND-CPA security. Our caching mechanisms are designed to be scheme-agnostic and can be applied to various FHE schemes including BFV, BGV, and CKKS, all of which base their security on the Ring Learning with Errors (RLWE) problem. We rely on additive re-randomization: if $c$ encrypts $m$, then $c \oplus \mathsf{Enc}_{pk}(0;\rho)$ is indistinguishable from a fresh encryption of $m$ for independently sampled $\rho$.

\textbf{Threat Model.}
We consider secure computation over encrypted bounded scalar values, including structured data (integers, categorical labels, scores) and quantized image data (8-bit intensities). The scenario involves outsourcing encrypted data to an untrusted server that performs computations without decryption keys. The adversary—either external or insider—has full access to ciphertexts and computational transcripts but no secret keys. While FHE enables computation on ciphertexts without exposing plaintext, it incurs significant overhead, particularly from bootstrapping. Our work optimizes encryption strategies for bounded scalar domains to minimize noise growth and reduce bootstrapping frequency, making FHE more practical while preserving security guarantees.
Each zero ciphertext in $\mathcal{Z}$ is independently generated, kept private by the encrypting client, used only once, and refreshed before pool exhaustion; thus, every reconstruction incorporates fresh randomness even when $\mathsf{Eval}$ is deterministic.

\textbf{Security of Parity-based Caching.} 
Note that this construction is compatible with BFV, BGV, and CKKS schemes as they all operate over similar ring structures and rely on RLWE for security. Let $\Pi(m)$ denote the encryption function of a plaintext $m$ under the base FHE scheme. For a plaintext $m = 2v + \delta$ with $v \in \{0,\dots,\lfloor M/2 \rfloor\}$ and $\delta \in \{0,1\}$, the cached encryption takes the form $\widetilde{\Pi}(m) = \Pi(2v) \oplus \delta \cdot \Pi(1) \oplus Z$, where $\Pi(2v)$ is from the precomputed even-cache, $\Pi(1)$ is cached, and $Z$ is a fresh encryption of $0$ sampled from the zero-pool $\mathcal{Z}$.

Assume for the sake of contradiction that the modified scheme $\widetilde{\Pi}$ with parity-based caching is not IND-CPA secure. Let $\mathrm{CPA}^{\mathcal{A}}_X$ denote the indistinguishability experiment with scheme $X$. By assumption, the base scheme is IND-CPA secure with $\Pr[\mathrm{CPA}^{\mathcal{A}}_{\Pi} = 1] \leq \frac{1}{2} + \varepsilon$, where $\varepsilon$ is negligible.

In $\widetilde{\Pi}$, each ciphertext is computed as the sum of cached $\Pi(2v)$, $\delta \cdot \Pi(1)$, and a fresh zero ciphertext $Z$. Each cached component $\Pi(2v)$ and $\Pi(1)$ is itself an RLWE sample indistinguishable from random, and adding an independently sampled $Z=\Pi(0;\rho)$ re-randomizes the ciphertext distribution. The determinism of $\mathsf{Eval}$ does not make $\widetilde{\Pi}$ deterministic, because the independently generated one-time ciphertext $Z$ introduces fresh randomness into every output. Therefore, the challenge ciphertext distribution under $\widetilde{\Pi}$ is computationally indistinguishable from that under $\Pi$:
{\small
\begin{equation}\label{eq:gap-poly}
\left|\Pr[\mathrm{CPA}^{\mathcal{A}}_{\widetilde{\Pi}} = 1] - \Pr[\mathrm{CPA}^{\mathcal{A}}_{\Pi} = 1]\right| \leq \mu(\lambda).
\end{equation}
}
Thus:
{\small
\begin{equation}\label{eq:final-parity}
\Pr[\mathrm{CPA}^{\mathcal{A}}_{\widetilde{\Pi}} = 1]
\leq
\frac{1}{2} + \varepsilon + \mu(\lambda).
\end{equation}
}
Since both $\varepsilon$ and $\mu(\lambda)$ are negligible, their sum is negligible. Therefore, the probability that $\mathcal{A}$ succeeds in the $\mathrm{CPA}^{\mathcal{A}}_{\widetilde{\Pi}}$ experiment is only negligibly higher than $\frac{1}{2}$, contradicting our assumption and proving that parity-based caching is IND-CPA secure.

\textbf{Security of Square-Root-based Caching.} Let $M$ be the maximum value in the bounded integer domain $\mathcal{D}=\{0,1,\dots,M\}$ (e.g., $M=255$ for 8-bit pixels). The construction is compatible with BFV, BGV, and CKKS, since these schemes share similar ring structures and rely on RLWE security. Let $\Pi(m)$ denote the encryption of plaintext $m$ under the base FHE scheme. For $v \in \mathcal{D}$, its cached encryption is
{\small
\begin{equation}
c = \widetilde{\Pi}(v) = \Pi(k) \odot \mathrm{pt}(k) \oplus \mathrm{pt}(r) \oplus Z,
\end{equation}
}
where $v = k^2 + r$, $k = \lfloor\sqrt{v}\rfloor$, $r = v - k^2 \in \{0,1,\dots,2\lfloor\sqrt{M}\rfloor\}$, and $Z$ is a fresh zero encryption from the zero-pool $\mathcal{Z}$. Assume for contradiction that the modified scheme $\widetilde{\Pi}$ with square-root-based caching is not IND-CPA secure. Let $\mathrm{CPA}^{\mathcal{A}}_X$ denote the indistinguishability experiment for scheme $X$. The success probabilities of $\mathcal{A}$ against $\Pi$ and $\widetilde{\Pi}$ are $\Pr[\mathrm{CPA}^{\mathcal{A}}_{\Pi} = 1]$ and $\Pr[\mathrm{CPA}^{\mathcal{A}}_{\widetilde{\Pi}} = 1]$, respectively. Here, $\lambda$ is the security parameter of the encryption scheme, determined by the underlying RLWE dimension and unrelated to the number of image pixels. By assumption, $\Pr[\mathrm{CPA}^{\mathcal{A}}_{\Pi} = 1] \leq \frac{1}{2} + \varepsilon$, where $\varepsilon$ is negligible.

In $\widetilde{\Pi}$, each ciphertext is formed from the cached $\Pi(k)$ by a public plaintext multiplication $\odot\,\mathrm{pt}(k)$, a plaintext addition $\oplus\,\mathrm{pt}(r)$, and a fresh zero $Z$. Each $\Pi(k)$ is itself an RLWE sample indistinguishable from random; the plaintext operations are public and, given $k$ and $r$, message-independent; and adding an independently sampled $Z=\Pi(0;\rho)$ re-randomizes the ciphertext distribution, so repeated encryptions of the same value differ with overwhelming probability. Therefore the challenge distribution under $\widetilde{\Pi}$ is computationally indistinguishable from that under $\Pi$, and $|\Pr[\mathrm{CPA}^{\mathcal{A}}_{\widetilde{\Pi}} = 1] - \Pr[\mathrm{CPA}^{\mathcal{A}}_{\Pi} = 1]| \leq \mu(\lambda)$. Combining the above,
{\small
\begin{equation}
\Pr[\mathrm{CPA}^{\mathcal{A}}_{\widetilde{\Pi}} = 1]
\leq
\frac{1}{2} + \varepsilon + \mu(\lambda).
\end{equation}
}
The last two terms are negligible: $\varepsilon$ is negligible by the IND-CPA security of the base scheme, and $\mu(\lambda)$ is negligible by the computational indistinguishability argument above. Hence, $\mathcal{A}$ succeeds in $\mathrm{CPA}^{\mathcal{A}}_{\widetilde{\Pi}}$ with probability only negligibly greater than $\frac{1}{2}$, contradicting the assumption. Therefore, the scheme with square-root-based caching is IND-CPA secure.

\section{Evaluation}
\label{evaluation}

This section presents our experimental setup (Section~\ref{experimentalSetup}), performance comparisons (Section~\ref{comparisonsection}), ablation studies (Section~\ref{ablation}), real-world applications (Section~\ref{casestudy}), bootstrapping frequency analysis (Section~\ref{sec:bootstrap_frequency}), and higher-bit-depth scalability (Section~\ref{sec:bitdepth}).

\begin{figure}[t]
  \centering
  \includegraphics[width=0.8\linewidth]{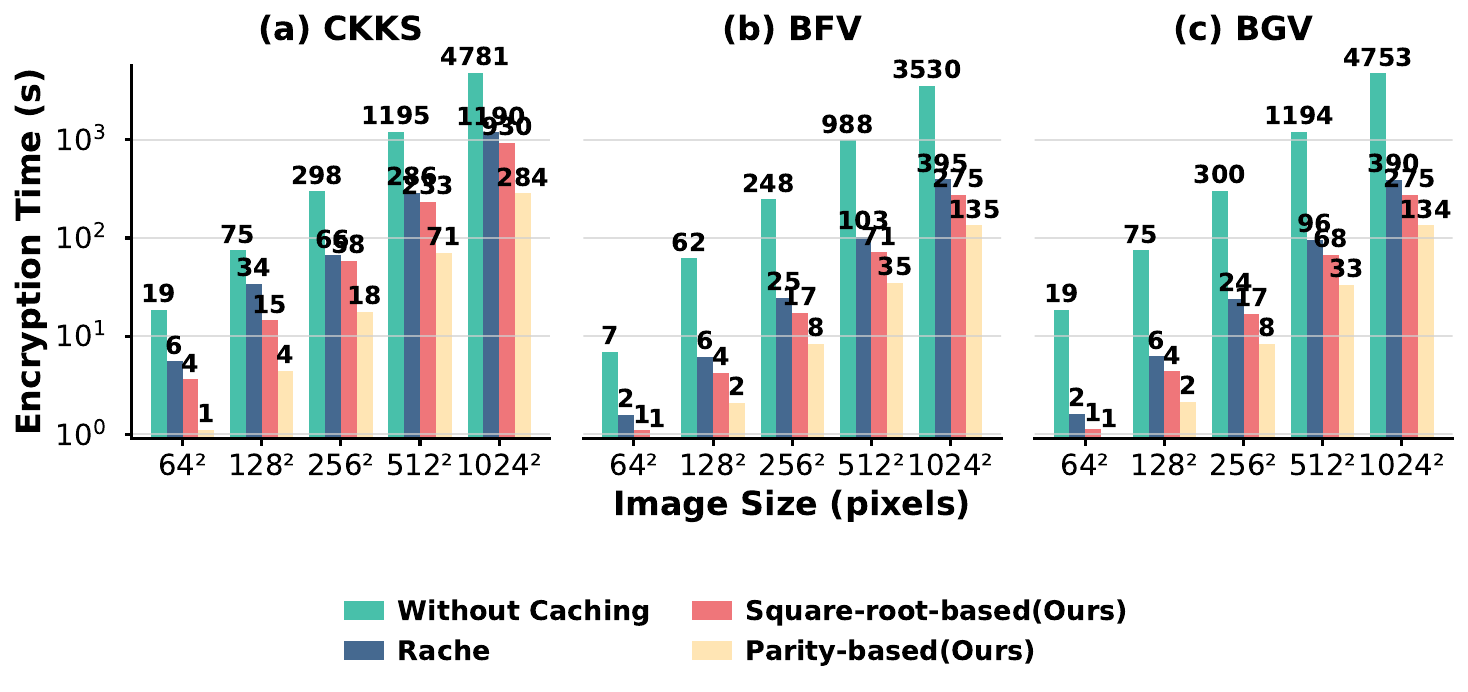}
  \caption{Comparison of encryption time for three schemes across different image sizes.}
  \label{fig:fhe-comparison}
\end{figure}

\begin{table}[t]
\setlength{\abovecaptionskip}{2pt}
\centering
\scriptsize

\caption{Comparison of encryption time(s) for three FHE schemes with varying $\log QP$.}
\label{tab:runtime_comparison}

\resizebox{\textwidth}{!}{%
\begin{tabular}{c c c ccccc c c ccccc}
\toprule

\multicolumn{3}{c}{} 
& \multicolumn{5}{c}{\textbf{logQP}}
& \multicolumn{2}{c}{} 
& \multicolumn{5}{c}{\textbf{logQP}} \\

\cmidrule(lr){4-8}
\cmidrule(lr){11-15}

\textbf{Scheme}
& \textbf{logN}
& \textbf{Method}
& 437 & 531 & 626 & 726 & 831
& \textbf{logN}
& \textbf{Method}
& 437 & 531 & 626 & 726 & 831 \\

\midrule

\multirow{8}{*}{CKKS}

& \multirow{4}{*}{12}
& Without Caching
& 18.62 & 28.71 & 38.56 & 48.98 & 60.55
& \multirow{4}{*}{14}
& Without Caching
& 79.53 & 125.71 & 172.31 & 220.76 & 275.68 \\

& & Rache
& 3.70 & 7.64 & 10.39 & 13.68 & 18.32
&
& Rache
& 13.06 & 27.94 & 45.11 & 70.62 & 89.25 \\

& & Sqrt-based (Ours)
& 3.63 & 5.59 & 7.50 & 9.53 & 11.77
&
& Sqrt-based (Ours)
& 15.47 & 24.46 & 33.54 & 42.95 & 53.58 \\

& & Parity-based (Ours)
& \textbf{1.11} & \textbf{1.71} & \textbf{2.29} & \textbf{2.93} & \textbf{3.60}
&
& Parity-based (Ours)
& \textbf{4.73} & \textbf{7.48} & \textbf{10.25} & \textbf{13.08} & \textbf{16.40} \\

\cmidrule(lr){2-8}
\cmidrule(lr){10-15}

& \multirow{4}{*}{13}
& Without Caching
& 38.37 & 59.81 & 81.19 & 103.17 & 127.05
& \multirow{4}{*}{15}
& Without Caching
& 260.38 & 346.95 & 369.96 & 492.74 & 622.45 \\

& & Rache
& 7.00 & 13.62 & 21.69 & 29.20 & 36.25
&
& Rache
& 28.09 & 72.08 & 115.29 & 156.60 & 195.77 \\

& & Sqrt-based (Ours)
& 7.46 & 11.63 & 15.80 & 20.09 & 24.74
&
& Sqrt-based (Ours)
& 50.67 & 67.54 & 71.97 & 95.98 & 121.05 \\

& & Parity-based (Ours)
& \textbf{2.28} & \textbf{3.46} & \textbf{4.69} & \textbf{5.96} & \textbf{7.34}
&
& Parity-based (Ours)
& \textbf{15.48} & \textbf{20.64} & \textbf{21.99} & \textbf{29.06} & \textbf{36.98} \\

\midrule

\multirow{8}{*}{BFV}

& \multirow{4}{*}{12}
& Without Caching
& 14.00 & 20.40 & 26.55 & 33.13 & 40.35
& \multirow{4}{*}{14}
& Without Caching
& 58.74 & 86.85 & 115.68 & 146.99 & 176.96 \\

& & Rache
& 1.58 & 2.93 & 4.48 & 5.71 & 7.16
&
& Rache
& 5.94 & 12.20 & 18.24 & 26.32 & 33.70 \\

& & Sqrt-based (Ours)
& 1.10 & 2.04 & 3.12 & 3.97 & 4.99
&
& Sqrt-based (Ours)
& 4.14 & 8.49 & 12.71 & 18.33 & 23.46 \\

& & Parity-based (Ours)
& \textbf{0.54} & \textbf{1.00} & \textbf{1.53} & \textbf{1.94} & \textbf{2.44}
&
& Parity-based (Ours)
& \textbf{2.03} & \textbf{4.16} & \textbf{6.22} & \textbf{8.97} & \textbf{11.48} \\

\cmidrule(lr){2-8}
\cmidrule(lr){10-15}

& \multirow{4}{*}{13}
& Without Caching
& 28.25 & 42.34 & 55.75 & 69.69 & 83.47
& \multirow{4}{*}{15}
& Without Caching
& 120.71 & 182.32 & 248.68 & 312.12 & 371.11 \\

& & Rache
& 2.78 & 5.80 & 8.77 & 11.79 & 15.57
&
& Rache
& 11.89 & 26.80 & 41.66 & 54.78 & 67.31 \\

& & Sqrt-based (Ours)
& 1.94 & 3.99 & 6.11 & 8.20 & 10.85
&
& Sqrt-based (Ours)
& 8.28 & 18.65 & 28.98 & 38.17 & 46.78 \\

& & Parity-based (Ours)
& \textbf{0.95} & \textbf{1.98} & \textbf{3.00} & \textbf{4.01} & \textbf{5.30}
&
& Parity-based (Ours)
& \textbf{4.05} & \textbf{9.12} & \textbf{14.18} & \textbf{18.63} & \textbf{22.93} \\

\midrule

\multirow{8}{*}{BGV}

& \multirow{4}{*}{12}
& Without Caching
& 17.01 & 26.86 & 36.77 & 46.85 & 57.49
& \multirow{4}{*}{14}
& Without Caching
& 73.34 & 116.55 & 160.50 & 205.56 & 256.51 \\

& & Rache
& 1.54 & 2.92 & 4.51 & 5.86 & 7.15
&
& Rache
& 5.85 & 11.92 & 18.60 & 25.88 & 33.63 \\

& & Sqrt-based (Ours)
& 1.08 & 2.05 & 3.17 & 4.12 & 5.02
&
& Sqrt-based (Ours)
& 4.11 & 8.38 & 13.07 & 18.20 & 23.64 \\

& & Parity-based (Ours)
& \textbf{0.53} & \textbf{1.01} & \textbf{1.55} & \textbf{2.02} & \textbf{2.46}
&
& Parity-based (Ours)
& \textbf{2.02} & \textbf{4.09} & \textbf{6.40} & \textbf{8.91} & \textbf{11.58} \\

\cmidrule(lr){2-8}
\cmidrule(lr){10-15}

& \multirow{4}{*}{13}
& Without Caching
& 35.33 & 56.06 & 76.79 & 98.16 & 120.20
& \multirow{4}{*}{15}
& Without Caching
& 151.69 & 245.38 & 338.29 & 439.94 & 536.78 \\

& & Rache
& 2.96 & 5.77 & 8.62 & 12.25 & 15.04
&
& Rache
& 11.46 & 26.02 & 42.11 & 54.66 & 69.26 \\

& & Sqrt-based (Ours)
& 2.08 & 4.05 & 6.06 & 8.62 & 10.57
&
& Sqrt-based (Ours)
& 8.06 & 18.33 & 29.57 & 38.91 & 49.13 \\

& & Parity-based (Ours)
& \textbf{1.02} & \textbf{1.99} & \textbf{2.97} & \textbf{4.02} & \textbf{5.17}
&
& Parity-based (Ours)
& \textbf{3.94} & \textbf{8.95} & \textbf{14.54} & \textbf{20.04} & \textbf{23.85} \\

\bottomrule
\end{tabular}%
}

\end{table}

\subsection{Experimental Setup}
\label{experimentalSetup}

Our implementation is based on the Lattigo library~\cite{lattigo}, an open-source Go library for lattice-based cryptography, and \thename\ contains about 10K lines of Go code. All experiments run on a CloudLab Clemson cluster with two 32-core AMD~7542 processors at 2.9 GHz, 512 GB RAM, and 2 TB SSD, using Ubuntu~20.04~LTS. We use the USC-SIPI Image Database~\cite{USCSIPI} and generate image subsets at resolutions 64×64, 128×128, 256×256, 512×512, and 1024×1024. The dataset includes multiple categories such as aerials, miscellaneous objects, sequences, and textures.

\subsection{Performance Comparison}
\label{comparisonsection}

\subsubsection{Time and Scalability}
\label{comparisontime}

We evaluate encryption efficiency across image sizes and schemes (CKKS, BFV, BGV) using the same parameters (\(\log N=12\), \(\log QP=109\); see Table~\ref{table:parameters}). Each test is run three times and we report the average. Figure~\ref{fig:fhe-comparison} shows encryption time on a logarithmic scale. The baseline without caching increases rapidly with image size for all schemes because each ciphertext requires fresh randomness and full polynomial generation. 
We use a per-value baseline because per-ciphertext cost is precisely what we reduce; SIMD packing cuts the number of encryptions, not the cost of each one. The parity-based caching method provides the best performance, achieving up to \(35\times\) speedup and consistently outperforming Rache and square-root-based caching across all schemes. Its efficiency comes from reusing pre-encrypted ciphertexts and updating only polynomial coefficients through lightweight homomorphic operations, reducing encryption to fast reconstruction with \(O(1)\) cache access.

\subsubsection{Parameter Impact on Performance}
\label{diff_params}

\begin{figure}[t]

  \centering
  \includegraphics[width=0.8\linewidth]{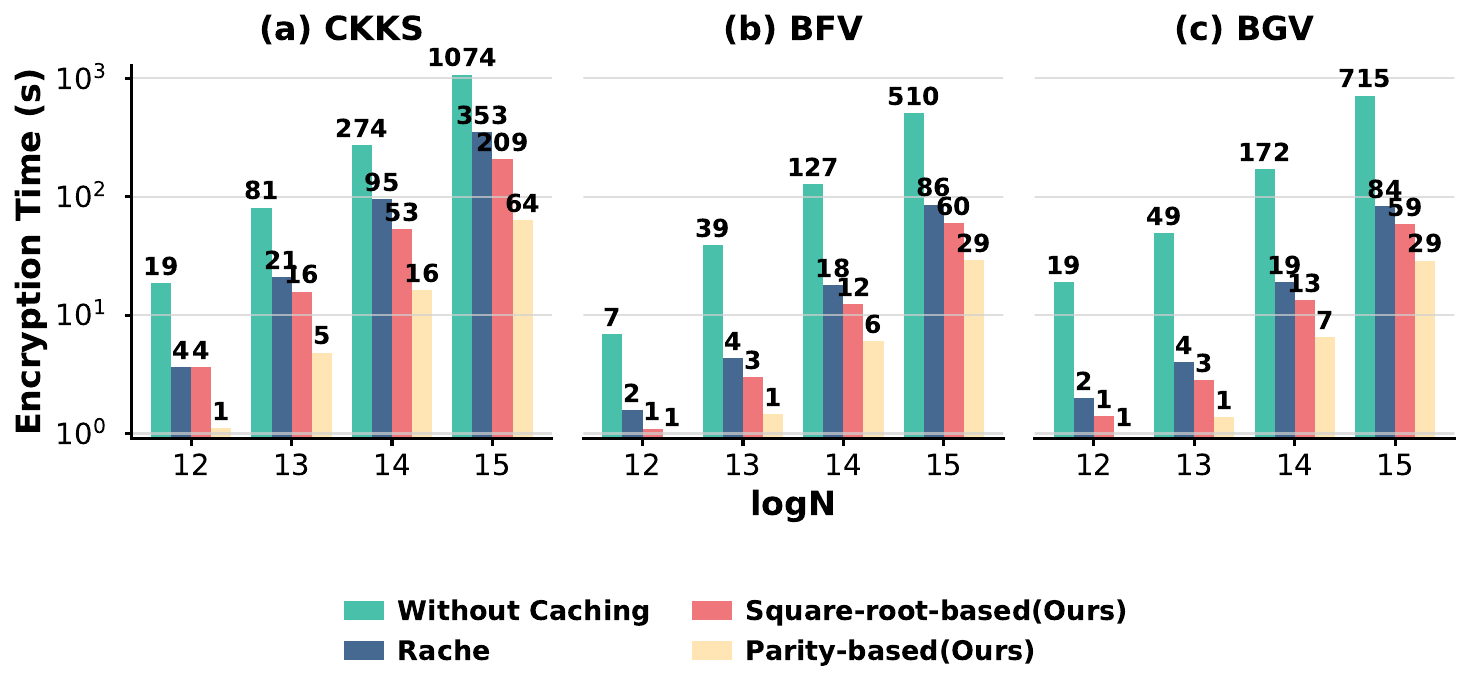}

  \caption{Comparison of encryption time for three FHE schemes with varying LogN.}
  \label{fig:3fhe_64x64}
\end{figure}

We evaluate the performance of our caching methods across CKKS, BFV, and BGV by varying \(\log N\) and \(\log QP\), comparing four approaches: without caching, Rache, square-root-based caching, and parity-based caching. As \(\log N\) increases (Figure~\ref{fig:3fhe_64x64}), encryption time for the baseline grows rapidly due to larger ciphertexts, while square-root-based caching reduces encryptions through decomposition and parity-based caching offers the best performance via \(O(1)\) ciphertext lookup. At the largest \(\log N\) values, parity-based caching achieves up to \(19\times\), \(18\times\), and \(25\times\) speedups in CKKS, BFV, and BGV, respectively.

When varying \(\log QP\) (Table~\ref{tab:runtime_comparison}), parity-based caching remains the fastest method across all evaluated configurations, while square-root-based caching requires less cache memory. Overall, both methods consistently accelerate encryption across schemes and parameters, and because most non-modulus parameters in Lattigo only marginally affect encryption time, these results highlight the generality and robustness of our approach.

\begin{table}[ht]
\setlength{\abovecaptionskip}{2pt}
\setlength{\tabcolsep}{2.7pt}
\renewcommand{\arraystretch}{0.8}
\centering
\footnotesize
\caption{Encryption time (s) across image types for four strategies. Parity-based achieves the lowest encryption time in all categories. The speedup is measured against the No Cache baseline.}
\begin{tabular}{lccccc}
\toprule
\textbf{Type} & \textbf{No Cache} & \textbf{Rache} & \textbf{Sqrt. (Ours)} & \textbf{Parity (Ours)} & \textbf{Speedup} \\
\midrule
Aerials   & 297 & 162 & 71 & \textbf{19} & 15.6$\times$ \\
Misc.     & 331 & 180 & 83 & \textbf{25} & 13.2$\times$ \\
Sequences & 311 & 170 & 68 & \textbf{18} & 17.3$\times$ \\
Textures  & 281 & 153 & 68 & \textbf{18} & 15.6$\times$ \\
\bottomrule
\end{tabular}
\label{tab:pic_type}
\end{table}

\subsubsection{Trade-off Analysis}
\label{tradeoff}

We summarize the trade-offs among precomputation, online time, and cache memory
using CKKS at \(\log N=12\) on \(64\times64\) images (trends hold at higher
resolutions). The no-caching baseline does no precomputation and is slowest
online (19 s). Rache~\cite{sigmod2023} adds 0.032 s precomputation and a 1 MB
cache, cutting online time to 6 s. Our methods shift more work offline: parity
uses 0.512 s and 16 MB to reach 1 s, while square-root uses 0.064 s and 2 MB to
reach 4 s.


\subsubsection{Impact of Image Categories}
\label{diff_imagetypes}

To evaluate performance across different image categories, we select 200 images of size \(256\times256\) from textures, aerial scenes, miscellaneous objects, and sequences in the USC-SIPI dataset~\cite{USCSIPI}, using CKKS with parity-based caching.

As shown in Table~\ref{tab:pic_type}, PixCrypt achieves speedups ranging from \(13.2\times\) to \(17.3\times\) across all four categories, indicating stable performance across different image content.



\begin{table}[t]
\setlength{\abovecaptionskip}{2pt}
\centering
\caption{Ablation study of the two layers in our method.}
\label{tab:ablation}
\resizebox{0.8\columnwidth}{!}{%
\begin{tabular}{llccc}
\toprule
\textbf{Scheme} & \textbf{Method} & \textbf{logN} & \textbf{Randomization Time (s)} & \textbf{Encryption Time (s)} \\
\midrule
\multirow{6}{*}{\textbf{CKKS}}
& \multirow{3}{*}{\textbf{Sqrt-based}}
& 12 & \textbf{3.74}   & \textbf{3.86}   \\
&  & 13 & 22.06  & 22.31  \\
&  & 14 & 110.4  & 111.14 \\
\cmidrule(lr){2-5}
& \multirow{3}{*}{\textbf{Parity-based}}
& 12 & \textbf{1.01}   & \textbf{1.01}   \\
&  & 13 & 6.39   & 6.39   \\
&  & 14 & 23.33  & 23.33  \\
\midrule
\multirow{6}{*}{\textbf{BFV}}
& \multirow{3}{*}{\textbf{Sqrt-based}}
& 12 & \textbf{3.90}   & \textbf{3.98}   \\
&  & 13 & 10.86  & 10.95  \\
&  & 14 & 52.49  & 52.62  \\
\cmidrule(lr){2-5}
& \multirow{3}{*}{\textbf{Parity-based}}
& 12 & \textbf{1.72}   & \textbf{1.72}   \\
&  & 13 & 5.03   & 5.03   \\
&  & 14 & 23.61  & 23.61  \\
\midrule
\multirow{6}{*}{\textbf{BGV}}
& \multirow{3}{*}{\textbf{Sqrt-based}}
& 12 & \textbf{3.87}   & \textbf{3.95}   \\
&  & 13 & 11.69  & 11.85  \\
&  & 14 & 46.78  & 46.92  \\
\cmidrule(lr){2-5}
& \multirow{3}{*}{\textbf{Parity-based}}
& 12 & \textbf{1.73}   & \textbf{1.73}   \\
&  & 13 & 4.73   & 4.73   \\
&  & 14 & 7.47   & 7.47   \\
\bottomrule
\end{tabular}%
}
\end{table}

\subsection{Ablation Studies}
\label{ablation}

To quantify the cost of randomization layer, we isolate ciphertext re-randomization time from the total encryption time across CKKS, BFV, and BGV under varying \(\log N\), as shown in Table~\ref{tab:ablation}. For parity-based caching, randomization time is virtually identical to  total encryption time across all schemes and $\log N$ settings (e.g., 1.01\,s 
vs.\ 1.01\,s for CKKS at $\log N{=}12$), showing that randomization accounts for most of the online reconstruction cost. For 
sqrt-based caching, a small gap exists (e.g., 3.74\,s vs.\ 3.86\,s for CKKS 
at $\log N{=}12$), attributable to the ciphertext--plaintext multiplication 
in the square decomposition rather than re-randomization itself.

These results show that randomization constitutes the primary encryption cost, which is necessary to ensure IND-CPA security, while other operations introduce negligible overhead.

\begin{figure}[t]

  \centering
  \includegraphics[width=0.8\linewidth]{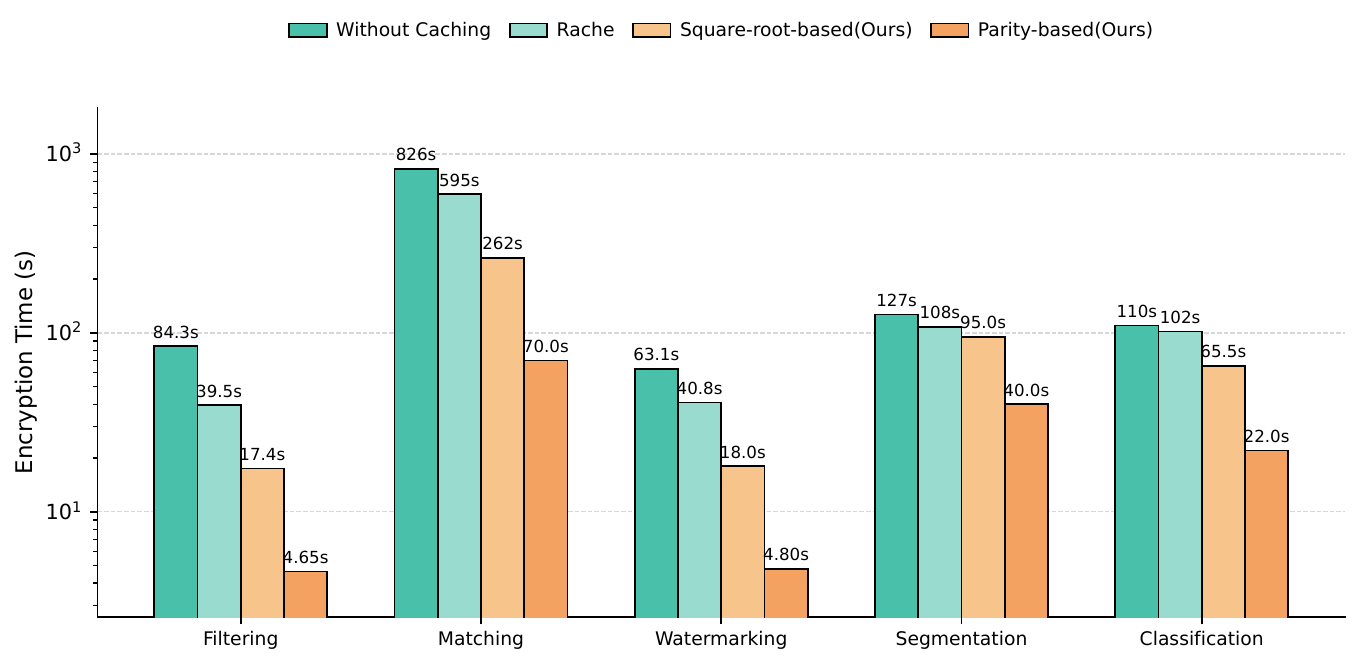}
  \caption{Encryption time (s) across real-world applications. PixCrypt consistently reduces runtime across all tasks, achieving up to 18× speedup.}
  \label{fig:applications}
\end{figure}

\subsection{Real-world Applications}
\label{casestudy}

To demonstrate that \thename~enables practical ciphertext-based image processing, 
we evaluate it on five representative tasks that require pixel-level operations, as shown in Figure~\ref{fig:applications}:

\textbf{Mean Filtering.}
We implement 3$\times$3 mean filtering~\cite{meanfilter} on 64$\times$64 encrypted images. As expected, the denoised output yields noticeable visual differences (MSE=218.13, PSNR=24.74). Compared to the 84.29s baseline, \thename~reduces encryption time to 4.65s, achieving 18× speedup, with similar gains for Gaussian filters.

\textbf{Image Matching.}
We compute encrypted L1 distances between 128$\times$128 images for similarity evaluation, enabling private image identification in contexts such as face recognition~\cite{lowe2004distinctive}. \thename~reduces encryption time by 12×, from 826.19s to 69.99s.

\textbf{Ciphertext-based Watermarking.}
To assert ownership over encrypted media~\cite{fridrich1998image}, we embed an imperceptible watermark by modifying a selected pixel (e.g., +5.0). Only differences above the threshold are detectable in a visual diff map. \thename~achieves a 13.1× speedup, reducing time from 63.08s to 4.8s.

\textbf{Image Segmentation.}
The encrypted image segmentation performs binary thresholding via a third-degree Taylor approximation of the tanh function to replace direct pixel comparisons. The proposed PixCrypt framework reduces encryption time for a 128×128 image by 3×, from 127 s to 40 s.

\textbf{Binary Classification.}
We implement binary classification on encrypted feature vectors using SIMD (Single Instruction, Multiple Data) batching. A compact CNN and quantized logistic head enable efficient encrypted inference, where each vector is packed into one ciphertext for linear scoring. This SIMD-aware design parallelizes computation and reduces runtime from 110 s to 22 s with accuracy comparable to plaintext.

\subsection{Bootstrapping Frequency Analysis}
\label{sec:bootstrap_frequency}
To empirically validate the noise reduction claim of
Section~\ref{bootstrapping}, we measure the bootstrapping frequency during
image segmentation, one of the deepest circuits in our evaluation. For a target
value $v$, the kernel approximates $\tanh(v)\approx v-\tfrac{v^3}{3}$, which
requires two sequential ciphertext--ciphertext multiplications and is therefore
sensitive to noise accumulation. We set the bootstrapping threshold to the
minimum ciphertext level $\ell_{\min}$ below which a bootstrap is triggered, and
evaluate all methods under identical CKKS parameters and the same $\ell_{\min}$.

Let $\sigma_0$ be the noise of a freshly constructed ciphertext, $\mathsf{thr}$
the noise threshold, and $\Delta_{\mathrm{add}},\Delta_{\mathrm{mult}}$ the
per-operation increments of plaintext--ciphertext and ciphertext--ciphertext
operations. The available budget is $B=\mathsf{thr}-\sigma_0$ and the
operational depth before a bootstrap is $d=\lfloor B/\Delta\rfloor$. A
reconstruction that stays in the additive, plaintext-oriented regime grows noise
linearly,
\begin{equation}
\mathrm{noise}(t)=\sigma_0+\sum_{j=1}^{t}\Delta_{\mathrm{add},j}+O(\log M)
=O\!\left(\sigma_0+t\,\Delta_{\mathrm{add}}+\log M\right),
\end{equation}
in contrast to the multiplicative accumulation
$O\!\left(\sigma_0\cdot\Delta_{\mathrm{mult}}^{\,t}\right)$ of chained
ciphertext--ciphertext construction, which yields
\begin{equation}
d_{\mathrm{add}}=O\!\left(\frac{\mathsf{thr}}{\Delta_{\mathrm{add}}}\right)
\;\gg\;
d_{\mathrm{mult}}=O\!\left(\log_{\Delta_{\mathrm{mult}}}\mathsf{thr}\right).
\end{equation}

A larger depth $d$ leaves fewer pixels crossing $\ell_{\min}$ within the fixed
segmentation circuit, hence fewer bootstraps. The two cached reconstructions differ in level consumption. Parity-based
synthesis is purely additive,
$\widetilde{\Pi}(t)=\Pi(2v)\oplus\delta\,\Pi(1)\oplus Z$, consuming no
multiplicative level and entering the circuit at the full budget $B$. Square-root
synthesis applies a ciphertext--plaintext multiplication
$\Pi(k)\odot\mathrm{pt}(k)$, which multiplies the scales
$\Delta_{ct}\!\cdot\!\Delta_{pt}$ and requires one rescaling
$\ell\!\to\!\ell-1$ to restore the target scale, lowering the budget by one
level. This predicts $d_{\mathrm{parity}}>d_{\mathrm{sqrt}}$, matching the
measured counts.

\begin{table*}
\centering
\caption{Normalized bootstrapping cost comparison under a constant per-bootstrap time assumption. We report the number of ciphertexts (pixels) requiring at least one bootstrap, along with the normalized bootstrapping time and total runtime.}
\label{tab:bootstrap_normalized}
\fontsize{9pt}{11pt}\selectfont
\renewcommand{\arraystretch}{1.2}

\begin{tabular}{lccccc}

\toprule
\textbf{Method} & \textbf{\#Bootstrap} & \textbf{BS Time (s)} & \textbf{Enc Time (s)} & \textbf{Total (s)} & \textbf{BS \%} \\
\midrule
Without Caching        & 64 & 164.463 & 0.931 & 165.394 & 99.4\% \\
Rache                 & 41 & 105.400 & 0.063 & 105.463 & 99.9\% \\
Sqrt-based (Ours)     & 34 & 87.400  & 0.287 & 87.687  & 99.7\% \\

Parity-based (Ours)   & 27 & 69.400  & 0.065 & 69.465  & 99.9\% \\

\bottomrule
\end{tabular}
\end{table*}

Table~\ref{tab:bootstrap_normalized} reports the number of bootstrapping
operations, total bootstrapping time, encryption time, and the runtime fraction
attributable to bootstrapping. Without caching, all $64$ pixels bootstrap,
confirming that independent fresh encryption offers no noise advantage. Rache
reduces this to $41$, but its ciphertext--ciphertext additions of independently
encrypted components keep the initial noise comparable to the baseline. The
square-root method reaches $34$ via plaintext--ciphertext multiplication, and the
level-preserving parity method attains the fewest at $27$, a $2.4\times$
reduction over the baseline. These counts follow the trend implied above:
the additive, level-preserving reconstruction sustains the largest depth $d$ and
thus the fewest bootstraps. Because bootstrapping accounts for over $99\%$ of runtime in this circuit, the
reduction in bootstrap count carries almost directly to the total time, which
drops from $165.4$\,s to $69.5$\,s, a $2.4\times$ end-to-end speedup spanning
encryption, homomorphic evaluation, and bootstrapping together, complementing
the encryption-time results of Section~\ref{evaluation}. The same table shows that encryption accounts for only a small fraction of the runtime in a deep circuit ($0.065$\,s out of $69.5$\,s). Therefore, the encryption speedups reported in Section~\ref{evaluation} are most relevant to shallow, encryption-heavy tasks such as filtering and watermarking, whereas deep circuits benefit primarily from fewer bootstrapping operations.


\subsection{Scalability to Higher Bit-Depth Domains}
\label{sec:bitdepth}
Our evaluation uses 8-bit pixels ($M=255$), but higher-precision imaging
(e.g., 12- or 16-bit medical images, $M$ up to $65{,}535$) raises a scalability
question. 
The parity and square-root caches store $\lfloor M/2\rfloor+1$ and
$\lfloor\sqrt{M}\rfloor+1$ ciphertexts, scaling as $O(M)$ and
$O(\sqrt{M})$, respectively. 
Table~\ref{tab:bitdepth} reports storage and precomputation across bit depths, based on $\approx125$\,KB per ciphertext and $\approx4$\,ms per encryption at $\log N=12$; its 8-bit results match our measurements. At 16-bit, parity-based caching needs $\sim$$4$\,GB and $\sim$$131$\,s, which is impractical, whereas square-root-based caching needs only $32$\,MB and
$\sim$$1$\,s and thus extends naturally to high precision. Parity can still be
used via a digit-wise split of a $w$-bit value into $\lceil w/8\rceil$ bytes over
a shared 8-bit cache, keeping storage at $O(256)$.
\begin{table}[tb]
\centering
\caption{Cache size and precomputation time of the two caching strategies across
plaintext bit depths. Parity scales as $O(M)$ and square-root as $O(\sqrt{M})$.}
\label{tab:bitdepth}
\fontsize{8pt}{9.5pt}\selectfont
\setlength{\tabcolsep}{4pt}
\renewcommand{\arraystretch}{1.05}
\begin{tabular}{llccc}
\toprule
\textbf{Bit depth ($M$)} & \textbf{Method} & \textbf{\#CTs} & \textbf{Cache} & \textbf{Precompute} \\
\midrule
\multirow{2}{*}{8-bit ($255$)}
 & Parity $O(M/2)$    & $128$   & $16$\,MB      & $0.51$\,s \\
 & Sqrt $O(\sqrt{M})$ & $16$    & $2$\,MB       & $0.06$\,s \\
\midrule
\multirow{2}{*}{12-bit ($4095$)}
 & Parity             & $2048$  & $256$\,MB     & $8.2$\,s \\
 & Sqrt               & $64$    & $8$\,MB       & $0.26$\,s \\
\midrule
\multirow{2}{*}{16-bit ($65535$)}
 & Parity             & $32768$ & $\sim4$\,GB   & $\sim131$\,s \\
 & Sqrt               & $256$   & $32$\,MB      & $\sim1.0$\,s \\
\bottomrule
\end{tabular}
\end{table}

\section{Conclusion}
\label{sec:conclusion}
This work introduces two orthogonal caching-based strategies to accelerate pixel-level fully homomorphic encryption (FHE): a \textit{square-root-based} method that decomposes values into square and remainder terms, and a \textit{parity-based} method that represents plaintexts as even bases plus parity bits. 
Both methods avoid ciphertext–ciphertext multiplications and reduce bootstrapping. We show that both methods preserve IND-CPA security under CKKS, BFV, and BGV, and experiments report up to \textbf{35$\times$} speedups per image.
\begin{credits}
\subsubsection{\ackname}
This work was supported by the National Science Foundation (NSF) under awards 2409851, 2528534, and 2403603, with additional partial support from NSF awards 1921576 and 2334196.
\end{credits}

%
%
%
\bibliographystyle{splncs04}

\bibliography{Reference}

\end{document}